\documentclass[a4paper,12pt]{article}

\usepackage{a4wide}
\usepackage[latin1]{inputenc}
\usepackage[english]{babel}
\usepackage{amssymb}
\usepackage{amsmath}
\usepackage{epsfig}

\newenvironment{entry}%
{\begin{list}{}{\setlength{\topsep}{0mm} \setlength{\itemsep}{0mm}
\setlength{\parskip}{0mm} \setlength{\parsep}{2mm}
\setlength{\leftmargin}{25mm} \setlength{\rightmargin}{4mm}
\setlength{\labelwidth}{20mm} \setlength{\labelsep}{2mm}}}%
{\end{list}}
\newcommand{\iteme}[1]{\item[\texttt{#1}\hfill]}

\newcommand{\ie}{{\it i.e.}}

\newcommand{\etc}{{\it etc.}}
\newcommand{\pythia}{{\sc Pythia}}
\newcommand{\herwig}{{\sc Herwig}}

\newcommand{\pt}[1]{p_{\perp, #1}}

\newcommand{\smallfrac}{\textstyle\frac}

\newcommand{\GeV}{\mathrm{\;GeV}}

\begin{document}

TSL/ISV-2005-0290\\
March 2005\\[5mm]

\begin{center}
{\Large MATCHIG: A program for matching charged Higgs boson production at
hadron colliders}\\
\bigskip\bigskip

{\large J.~Alwall}\\[2mm]
{High Energy Physics, Uppsala University, Box 535, S-75121 Uppsala, Sweden}\\[1mm]
E-mail: {\tt johan.alwall@tsl.uu.se}
\end{center}

\begin{abstract}
\noindent This manual describes how to use the MatCHig code for matching the
charged Higgs boson production processes $gg\to tbH^\pm$ and $gb\to
tH^\pm$. A negative term, correcting for the double-counting between
these processes, is implemented as an external process to \pythia,
allowing the Monte Carlo generation of matched events. Results from
the matching were published in
\cite{Alwall:2004xw}. The code can be downloaded from {\tt
http://www.isv.uu.se/thep/MC/matchig/} .
\end{abstract}

\section{Physics motivation}
Many extensions of the Standard Model, most notably supersymmetric
theories like the MSSM, predict the existence of a charged Higgs
boson. Since there is no charged scalar particle in the Standard
Model, the discovery of such a particle would be a clear signal for
physics beyond the Standard Model, and would provide insight into
the nature of the extension needed. However, in order to find the
charged Higgs we need an accurate description of its production and
what experimental signatures to look for.

Charged Higgs production at hadron colliders
is usually described in Monte Carlo event generators such as \pythia\ 
\cite{pythia} and \herwig\ \cite{herwig} using the $2\to2$ process 
\begin{equation}
\label{eq:LO}
gb\to tH^- \; (g\bar b\to \bar t H^+)\;.
\end{equation}
Here, the incoming $b$-quark is described by the DGLAP evolution
equations using gluon splitting to $b\bar b$ in the collinear
approximation. Therefore the event also contains an outgoing $\bar b$
($b$) quark. For low transverse momenta of this accompanying $b$
quark, the $2\to2$ process \eqref{eq:LO}, including initial state
parton showers, describes the cross-section well, since the $b$ quark
density includes a resummation of potentially large logarithms of the
type $\left(\alpha_s\log\frac{\mu_F}{m_b}\right)^n$, where $\mu_F$ is
the factorization scale. However, for large $\pt{b}$ (transverse
momentum of the accompanying $b$-quark) the collinear approximation is
no longer appropriate, and one need instead use the exact matrix
element of the $2\to3$ process
\begin{equation}
\label{eq:2to3}
gg\to t\bar b H^-\; (gg\to \bar t b H^+)\;.
\end{equation}

At low transverse momenta, the $2\to3$ process can be described in
terms of gluon splitting to $b\bar b$ (at order
$\alpha_s\log\frac{\mu_F}{m_b}$), times the matrix element of the
$2\to2$ process. This means that we can not simply add the two
processes without getting double-counting for low values of $\pt{b}$. This
problem is usually addressed by using only the $2\to2$ process
\eqref{eq:LO} when the accompanying $b$ quark is not observed, and
only the $2\to3$ process \eqref{eq:2to3} when observing the
accompanying $b$ quark. However, as was shown in
\cite{Alwall:2004xw}, for low transverse momenta ($\pt{b}\lesssim
100\GeV$) the latter approach underestimates the differential
cross-section. Therefore, when the accompanying $b$-quark is observed,
it is necessary to use both the $2\to2$ and the $2\to3$ processes
together, appropriately matched to remove the double-counting.

Detailed discussions, derivations, references and results from this
matching were published in \cite{Alwall:2004xw}.

\section{The double-counting term}
In order to remove the double-counting, we define a double-counting
term $\sigma_\mathrm{DC}$, given by the part of the $2\to3$ process
(eq.~\eqref{eq:2to3}) which is already included in the $2\to2$ process
(eq.~\eqref{eq:LO}). This term is then subtracted from the sum of the
cross-sections from the two processes,
\begin{equation} 
\sigma = \sigma_{2\to 2} +\sigma_{2\to 3} \, -\sigma_\mathrm{DC}\;.
\label{eq:xsec}
\end{equation}

The double-counting term is given by the leading ($\mathcal
O(\alpha_s\log\frac{\mu_F}{m_b})$) contribution of the $b$
quark density to the $2\to2$ process cross-section:
\begin{equation}
\label{eq:DC}
\sigma_\mathrm{DC}=\int dx_1dx_2\left[g(x_1,\mu_F)b'(x_2,\mu_F)
\frac{d\hat{\sigma}_{2\to 2}}{dx_1dx_2}(x_1,x_2) 
+ x_1 \leftrightarrow x_2\right]
\end{equation}
where $b'(x,\mu_F^2)$ is the leading order $b$-quark density given by
\begin{equation}
 b^\prime(x, \mu_F^2)\approx
 \frac{\alpha_s}{2\pi}\log\frac{\mu_F^2}{m_b^2}\int
 \frac{dz}{z} P_{qg}(z) \; g\left(\frac{x}{z},\mu_F^2\right)
\end{equation}
with $P_{qg}$ the $g\to q\bar q$ splitting function, $g(x,\mu_F^2)$
the gluon density function, $\mu_F$ the factorization scale and $z$
the longitudinal gluon momentum fraction taken by the $b$-quark.

Including kinematic constraints due to finite center of mass energy
and finite $b$ quark mass, the resulting expression for the
double-counting term can be written as
\begin{eqnarray} 
\lefteqn{\sigma_\mathrm{DC} = \int_{\tau_\mathrm{min}}^1\frac{d\tau}{\tau}
  \int_{\frac{1}{2}\log\tau}^{-\frac{1}{2}\log\tau}dy^*\frac{\pi}{\hat s}
  \int_{-1}^1\frac{\beta_{34}}{2}d(\cos\hat\theta)
  \left|\cal M_\mathrm{2\to2}\right|^2 \,\frac{\alpha_s(\mu_R^2)}{2\pi}\times}
  \nonumber \\
&&\left[\int_{x_1}^{z_\mathrm{max}}dz P_{qg}(z)
 \int_{Q^2_\mathrm{min}}^{Q^2_\mathrm{max}}\frac{d(Q^2)}{Q^2+m_b^2} 
  \, \frac{x_1}{z} g\left(\frac{x_1}{z},\mu_F^2\right) \,x_2 g(x_2,\mu_F^2) 
  + x_1 \leftrightarrow x_2\right] \;.
\label{eq:diff_dc}
\end{eqnarray}

Here $\cal M_\mathrm{2\to2}$ is the matrix element for the $2\to2$
process (\ref{eq:LO}), $\mu_F$ and $\mu_R$ are the factorization and
renormalization scales as in the $2\to3$ process, and the kinematical
variables are $\tau= x_1 x_2$, $x_{1,2}=\sqrt{\tau}e^{\pm y^*}$, $\hat
s = \tau s$. $\hat\theta$~is the polar angle of the $t$-quark in the
CM system of the $2\to 2$ scattering, and $\beta_{34} = \hat
s^{-1}\sqrt{(\hat s-m_t^2-m_{H^\pm}^2)^2-4m_t^2m_{H^\pm}^2}$. $Q^2$ is
the virtuality of the incoming $b$-quark and $z$ is identified with
the ratio of the center-of-mass energies of the $gb$ ($2\to2$) system
and the $gg$ ($2\to3$) system.

The integration limits are given by
\begin{subequations}
\label{eq:limits}
\begin{align}
\tau_\mathrm{min}& = (m_t+m_{H^\pm})^2/s\\
z_\mathrm{max} &= \frac{Q^2_\mathrm{opt} \hat s}
		{(Q^2_\mathrm{opt}+\hat s)(Q^2_\mathrm{opt} + m_b^2)}
 \;,\;Q^2_\mathrm{opt}=\min\left(\sqrt{\hat s m_b^2},\mu_{F,2\to2}^2\right)\\
Q^2_\mathrm{min}& = \smallfrac{1}{2}
		\left[\hat s\left(z^{-1}-1\right)-m_b^2\right] - 
	\smallfrac{1}{2}\sqrt{\left[\hat s\left(z^{-1}-1\right)-
		m_b^2\right]^2-4\hat sm_b^2}\\
Q^2_\mathrm{max}& = \min\left\{\mu_{F,2\to2}^2,\smallfrac{1}{2}
		\left[\hat s\left(z^{-1}-1\right)-m_b^2\right] + 
	\smallfrac{1}{2}\sqrt{\left[\hat s\left(z^{-1}-1\right)-
		m_b^2\right]^2-4\hat sm_b^2}\right\}
\end{align}
\end{subequations}
where $\mu_{F,2\to2}^2$ is the factorization scale of the $2\to2$
process, which sets the upper $Q^2$ limit in the parton showers.

\section{Implementation}

The double-counting term is implemented as an external process using
the Les Houches generic user process interface for event generators
\cite{Boos:2001cv}. This means that it can in principle be used
together with any event generator supporting the Les Houches standard,
but it is primarily intended for use with \pythia\ \cite{pythia}, and
it uses several \pythia\ routines and common blocks. The relevant
FORTRAN-files can be downloaded from \\
{\tt http://www.isv.uu.se/thep/MC/matchig/} .

The implementation, found in the file {\tt matchig.f}, provides the
subroutines {\tt UPINIT}, to set up the external process, and {\tt
UPEVNT}, which supplies an event (\ie\ a set of incoming and outgoing
particles and momenta) and a weight for the event. Please note that
since the double-counting contribution should be subtracted from the
sum of the positive processes \eqref{eq:LO} and
\eqref{eq:2to3}, this weight is negative for double-counting events. 
This means that if all three processes are run simultaneously in
\pythia, the total cross-section will be the correctly matched one given by
eq.~\eqref{eq:xsec}.

The events from the double-counting term should likewise be subtracted
in the analysis. In practice, this means that these events should
undergo the same cuts and the same detector response simulation, \etc,
as the events from the positive contributions, and then be added with
weight $-1$ to any histograms. The events from the double-counting
term can be identified in \pythia\ using the parameter construction
{\tt KFPR(MSTI(1),2)}, which contains the external process identifier
for external processes. The external process identifier for the
double-counting process is {\tt LPRUP(1)=10000}.

Since the $2\to2$ process has larger cross-section than the
double-counting term in all phase-space
regions, the risk of getting a negative total cross-section in any
phase-space point should be small given enough statistics.

\subsection{Parameters and recommended settings}

For information on how to use \pythia\ with external processes, please
see the \pythia\ manual \cite{pythia}, section 9.9. An example main
program, {\tt matchex.f}, illustrating how to use the double-counting
term together with \pythia, can also be found on the homepage. Please
note that the double-counting term should be run together with the
internal processes $gb\to tH^\pm$ ({\tt ISUB=161}) and $gg\to tbH^\pm$
({\tt ISUB=401}) to get matched events. (Note that the latter process
({\tt ISUB=401}) was implemented in \pythia\ from version 6.223.)

The following \pythia\ parameters are used in {\tt matchig.f}:\\
\begin{entry}
\iteme{PMAS} (in the common block {\sc pydat2}) is used to get masses and
widths for the charged Higgs boson ({\tt PMAS(37,1), PMAS(37,2)}), top
quark ({\tt PMAS(6,1), PMAS(6,2)}), bottom quark ({\tt PMAS(5,1)}) and
$W^\pm$ boson ({\tt PMAS(24,1)}).

\iteme{PARU(141)} (in {\sc pydat1}) is used to get the value of $\tan\beta$.

\iteme{PARU(102)} (in {\sc pydat1}) is used to get the value of 
$\sin^2\theta_W$.

\iteme{MSTP(39)} (in {\sc pypars}) is used to choose factorization and
renormalization scales as for the $2\to3$ process. The following
choices are possible: {\tt MSTP(39) = 3}, {\tt 5}, {\tt 6} or {\tt
8}. The options {\tt MSTP(39) = 6} and {\tt 8} are new from \pythia\
6.226, please see \cite{pyupdate} for details. Our recommended setting
is {\tt MSTP(39) = 8}, corresponding to {\tt MSTP(32) = 12}. Note that
in this case {\tt PARP(193)} and {\tt PARP(194)} must also be set, see
below.

\iteme{MSTP(32)} (in {\sc pypars}) is used to choose the maximum scale
for parton showers, \ie\ the maximum transverse momentum of the
accompanying $b$-quark as for the $2\to2$ process ($\mu_{F,2\to2}$ in
eq.~\eqref{eq:limits}). The following choices are possible: {\tt
MSTP(32) = 4}, {\tt 11} or {\tt 12}. The options {\tt MSTP(32) = 11}
and {\tt 12} are new from \pythia\ 6.226, please see \cite{pyupdate}
for details. Our recommended setting is {\tt MSTP(32) = 12},
corresponding to {\tt MSTP(39) = 8}. Note that in this case {\tt
PARP(193)} and {\tt PARP(194)} must also be set, see below.

\iteme{PARP(34)} (in {\sc pypars}) is used to modify the factorization
and renormalization scales as in \pythia. Note however that it does
not affect the maximum transverse momentum of the accompanying
$b$-quark.

\iteme{PARP(67)} (in {\sc pypars}) is used to modify the maximum scale
for parton showers, \ie\ the maximum transverse momentum of the
accompanying $b$-quark ($\mu_{F,2\to2}$ in eq.~\eqref{eq:limits}), as
in \pythia.

\iteme{PARP(48)} (in {\sc pypars}) is used to cut the tails of the 
Breit-Wigner distribution for the charged Higgs mass at masses outside
$m_{H^\pm}\pm\mathtt{PARP(48)}\!\cdot\!\Gamma_{H^\pm}$, as in \pythia.
\end{entry}

The values of $\alpha_s$, $\alpha_{em}$, the parton distributions and
the running top and bottom quark masses are calculated using the
corresponding \pythia\ subroutines.  The top quark and charged Higgs
masses are Breit-Wigner distributed with mass-dependent widths in the
same way as in the internal $gb\to tH^\pm$ and $gg\to tbH^\pm$
processes, using the \pythia\ subroutine {\tt PYWIDT}. Suppression of
the cross-section due to closed decay channels of the charged Higgs
boson and top quark is done as in the \pythia\ processes.

{\bf Factorization scale:} As shown in \cite{Alwall:2004xw}, there are
strong arguments that a non-standard factorization scale $\mu_F\approx
0.5\frac{m_{H^\pm}+m_t}{2}$ should be used in charged Higgs
production. The option to have fixed factorization and renormalization
scales of choice was added to \pythia\ from version 6.226
\cite{pyupdate}. This is done by setting {\tt MSTP(32)=12} (for
$2\to2$ processes) and {\tt MSTP(39)=8} (for $2\to3$ processes). Then
one needs to set {\tt PARP(193)} to the square of the factorization
scale ($\mu_F^2$), and {\tt PARP(194)} to the square of the
renormalization scale $\mu_R^2$ (usually taken to be $m_{H^\pm}^2$).

{\bf On-shell parton showers:} In order to be able to compare the
differential cross-section for the $2\to2$ process
\eqref{eq:LO}, the $2\to3$ process \eqref{eq:2to3} and the
double-counting term on parton level, one should use on-shell
initial-state parton showers, set by {\tt MSTP(63)=0}. In a realistic
case, using jet-finding algorithms, there should however be no
observable difference compared to having the parton showers off-shell.

\section{Final comments}
The code described in this document was written for research conducted
together with Johan Rathsman. If you use this code, please cite 
\cite{Alwall:2004xw}.

\end{document}